\documentclass[twocolumn,epjc3]{svjour3}
\usepackage{amsmath}
	  \usepackage{amssymb}
	 \usepackage{graphics}
	  \usepackage{epsfig}
	  \usepackage{hyperref}

\DeclareMathOperator{\Df}{{\mathcal{ D}}}
\DeclareMathOperator{\lagr}{\mathcal{ L}}

\begin{document}
\title{Quantum corrections to the Classical Statistical
 Approximation for the expanding quantum field}
\author{A.V. Leonidov \thanksref{addr1,addr2} \and
A.A. Radovskaya \thanksref{e1,addr1}}

\thankstext{e1}{e-mail: raan@lpi.ru}

\institute{  P.N.Lebedev Physical Institute, 119991, Moscow, Russia
\label{addr1}
\and
 Moscow Institute of Physics and Technology,
 Dolgoprudny, 141700, Moscow region, Russia
 \label{addr2}}

\date{Received: date / Accepted: date}

\maketitle

\begin{abstract}
	We found the deviation of the equation of state from
	ultrarelativistic one due to quantum corrections
	 for a nonequilibrium longitudinally expanding scalar field.
	Relaxation of highly excited quantum field is usually described in terms
	 of Classical Statistical Approximation (CSA). However, the expansion
	  of the system reduces the applicability
	  of such a semiclassical approach as the CSA
	  making quantum corrections important. We  calculate
	  the evolution of the trace of the energy-momentum tensor
	    within the Keldysh-Schwinger framework for
	   static and longitudinal expanding geometries. We provide
	   analytical and numerical arguments for the appearance
	   of the nontrivial intermediate regime
	   where quantum corrections are significant.
\end{abstract}

\section{Introduction}

Highly nonstationary dense quantum fields define
the initial stage of many physical problems.
These include physics of the early stage of
ultrarelativistic heavy ion collisions
\cite{G2015,K2016}, cold atomic gases \cite{berg1,berges_gases,Lee}
 and the processes in the early Universe \cite{cosm_1,cosm_2,cosm_3}.
 Theoretical description of such dense fields characterised
by high occupation numbers can be naturally based on
the classical approximation (classical solutions of the
equations of motion). In the physics of heavy
ion collision the corresponding solution is termed glasma \cite{LM2006}.

Quantitative description of the evolution of highly
  excited matter should %necessarily
  include resummation
        of the leading order (LO) quantum corrections.
        This is due to characteristic instabilities of tree-level
        dynamics that can appear in the form of the family of
        parametric resonances. The corresponding instabilities of glasma
        were first described in \cite{RR2006}. To overcome this problem one
        needs to resum the contributions of the corresponding quantum
        fluctuations. Such resummation demonstrates
        that the result can be rewritten in the form of averaging over
        classical trajectories with a distribution of the
        initial conditions. We refer to such an equivalence as to
        Classical Statistical Approximation (CSA).   To our knowledge
        this approximation for quantum field theory was
        introduced in the work  \cite{MM}, and the first diagrammatic
        proof of the equivalence of such resummation was given in works
        \cite{cosm_1}. In the literature on physics of the early stage of
        heavy ion collisions, this statement was proven and used in the analysis of
        quantum corrections to the evolution of glasma in \cite{FGM2007}.
        The present research is relevant to the profound works on aforementioned
        equivalence  used in a study of quantum corrections to the evolution of
        strong scalar field in static \cite{initial1,initial2} and expanding \cite{initial3} geometries.

The Keldysh-Schwinger (KS) technique (closed-time path formalism)
\cite{sch,kel} provides a systematic way of studying time-dependent
nonequilibrium phenomena in quantum field theory, see the recent
review in \cite{berg1}.  Within this formalism,  the CSA
(averaging over classical trajectories with different initial conditions)
does naturally arise at the leading order of the semiclassical
approximation \cite{chem1,chem2}.
 In the quantum field
theory context this was discussed in \cite{method} where the JIMWLK
equations \cite{JKLW1998,ILM2001a,FILM2002,ILM2001b} were shown to
follow from such a semiclassical expansion.
For the scalar field
model of \cite{initial1} such equivalence was established in \cite{LR1}.

An evident question of computing the quantum
corrections to the results of LO resummation/semiclassical
approximation is being risen.  Such NLO corrections to the LO resummation
of the evolving scalar field \cite{initial1,initial2} was
discussed in \cite{EGW2014} with a very thought-provoking
conclusion of their non-renormalizability. The computation of
the NLO corrections to the JIMWLK equations was described in
\cite{KLM2014}.

The fact that resummation of one-loop corrections results in LO term of the semiclassical approximation indicates that we are dealing not with a plain small coupling expansion.   For the CSA the initial conditions become rather important, in particular, the scale characterizing the initial field.
Computation of quantum corrections to the semiclassical approximation
in the KS formalism was discussed in \cite{BG2007} for the cold quantum gas.
A problem of computing NLO corrections to the evolution of quantum scalar
field in the model of \cite{initial1} was discussed in the two preceding
works \cite{LR1,LR2}. In the first work we described the systematic procedure
of computing quantum corrections in the framework of KS formalism,
derived  analytical expressions for pressure relaxation
in the scalar field model and wrote down explicit expressions
for the NLO corrections for one-point and two-point correlation functions.
In the second paper \cite{LR2} we derived analytical expressions
for the mean field, energy and pressure of the homogeneous scalar
field in the static geometry and discussed  the critical role of the
character of initial conditions for applicability of the CSA approximation.

In the present paper we study the NLO corrections to the evolution of the trace of energy-momentum tensor of the homogeneous scalar field in the static and expanding geometries. This problem is of particular interest for the physics of the early stage of heavy ion collision because the behavior of this trace is of direct relation to the issue of thermalization and isotropization of the initially produced highly excited matter \cite{G2015,K2016}, see the recent advanced analysis of this issue in \cite{KWI,KWII}.

The paper is organized as follows:

In section  \ref{sec:model} we describe the model under consideration
and discuss assumptions and simplifications which make derivation
of the analytical answers possible.

Section \ref{sec:static}  is devoted to the static geometry.
We calculate NLO corrections to the evolution of the trace of
the energy-momentum tensor and demonstrate that these corrections
do vanish at large times.

In  section \ref{sec:expanding} we perform calculations analogous to ones of
 section  \ref{sec:static}  but for the expanding geometry.
We conclude with the analytical prove of the existence of
the intermediate quasistationary regime with the equation
of state different from relativistic one.

In  section \ref{sec:numerical} we demonstrate the results
of the numerical calculations.

In  section \ref{sec:results} we summarise obtained
results and discuss the region of applicability of the CSA.

In   \ref{sec:app_nlo} we describe a general scheme
suitable for derivation of the quantum corrections to the CSA for the scalar
field theory $\varphi^4$.

\section{\label{sec:model}Model and assumptions}

The main object of our study is an evolution of the energy-momentum tensor of the highly excited quantum field in the massless scalar $\varphi^4$ theory
\begin{gather}
    \mathcal{L}=\frac{1}{2}\partial_{\mu}\varphi\partial^{\mu}\varphi
     - \frac{g^2}{4}\varphi^4 +J\varphi ,
     \label{4scalar}
\end{gather}
where the source $J$ is used for the construction of diagrammatic expansion only and is set to zero in all final expressions. This is the stylized model proposed to study the dynamics of nonequilibrium matter created at the early stages of heavy ion collisions in \cite{initial1}.
The observable that we are interested in is the canonical energy-momentum tensor
 \begin{gather}
     T^{\mu\nu} = \partial^{\mu}\varphi\partial^{\nu}\varphi
		 - g^{\mu\nu}\mathcal{L}.
  \end{gather}
 Of particular interest is the trace of the energy-momentum tensor including contributions of energy density and pressure. An existence of the definite relation between energy density and pressure (equation of state, EOS) is known to be a crucial prerequisite for hydrodynamic description of the problem under study.   For the homogeneous case ($\partial_{\bf x}\varphi = 0$)    the expressions for energy density $\varepsilon$ and pressure $p$ read
 \begin{gather}
  T^{\mu}_{\mu}=  \varepsilon - 3p,\nonumber\\
	\varepsilon = \frac{\dot\varphi^2}{2}+\frac{g^2 \varphi^4}{4},\nonumber\\
	\qquad p=\frac{\dot\varphi^2}{2}-\frac{g^2 \varphi^4}{4}.
	\label{trace}
 \end{gather}

At the classical level $T_{\mu}^{\mu}$   is a  periodic function \cite{initial1} and, therefore, the equation of state in this approximation does not exist. Summation of quantum corrections in the CSA approximation \cite{initial1,initial2} lead however to $\langle T_{\mu}^{\mu} \rangle =0$ and, therefore, to the EOS $\varepsilon = 3 \langle p \rangle$ expected for the ultrarelativistic liquid. In the present paper we continue the study of the quantum corrections to CSA began in \cite{LR1,LR2} with a particular focus on the case of expanding geometry.

\section{\label{sec:static}Static geometry}

In this section we consider the evolution of the energy-momentum tensor in the static  geometry.  The  action for the homogeneous scalar field theory under consideration reads
\begin{gather} \label{stat_lagr}
	S_{st} = V_3\int dt \Big( \frac{1}{2}\dot\varphi^2 - \frac{g^2}{4}\varphi^4 +
	J\varphi \Big), \quad
	V_3 = \int d^3x.
\end{gather}
The corresponding equation of motion
\begin{gather}
 \ddot \varphi + g^2 \varphi^3 = J
 \label{EoM}
\end{gather}
can be solved analytically for $J=0$ \cite{initial1} in terms of the Jacobi elliptical function $cn$ with module $k^2 = \frac{1}{2}$
\begin{gather}
 \phi_{cl}(t)=\phi_m cn\left(\frac{1}{2}, g\phi_m t + C\right)
 \label{class_sol}
\end{gather}
with the period $T_{cl}=\frac{4}{g \phi_m}K(1/2)$, where $K(1/2)$  is the complete elliptic integral of the first kind. The constants $\phi_m$ and $C$ are the amplitude and the phase of the solution.

The corresponding energy-momentum tensor reads
\begin{gather}
	T^{\mu\nu} = \mbox{diag}(\varepsilon,p,p,p),
\end{gather}
where the energy density and the pressure are given by eq.(\ref{trace}). The expression for its trace takes the form
\begin{gather}
	T^{\mu}_{\mu} = \varepsilon - 3p = -\dot\varphi^2 + g^2 \varphi^4.
\end{gather}

At  the classical level the trace
\begin{gather}
		\big(T^{\mu}_{\mu}\big)_{cl} = -\dot\phi^2_{cl} + g^2 \phi^4_{cl}
\end{gather}
is the function of the periodic classical solution (\ref{class_sol}), therefore the exact correspondence between the energy density and pressure is missing and it is necessary to study quantum evolution \cite{initial1}.

Temporal evolution of the energy-momentum tensor in the KS formalism from some initial state at $t=t_0$  till $t=t_1 $ is given by
\begin{gather}
	\langle T^{\mu}_{\mu} (\hat\varphi)\rangle_{t_1} =
	 \int d\xi\ D[\xi_1,\rho_0,\xi_2]\	 T^{\mu}_{\mu}(\xi) \times\\
		\int\limits_{\eta_F(t_0)=\xi_1}^{\eta_F(t_1)
	  =\xi} \mathcal{ D}\eta_F(t) \int\limits_{\eta_B(t_0)=\xi_2}^{\eta_B(t_1)
	 =\xi} \mathcal{ D} \eta_B(t)\ e^{iS_K[\eta_F,\eta_B]},\nonumber
\end{gather}
where $\hat\rho(t_0)$ is the density matrix of the initial field configuration,
\begin{gather}
	D[\xi_1,\rho_0,\xi_2] =  \int d\xi_1 \int d\xi_2\
	\langle\xi_1|\hat\rho(t_0)|\xi_2\rangle
\end{gather}
the Keldysh action is $S_K[\eta_F,\eta_B] =S[\eta_F]-S[\eta_B]$ and the fields
 $\eta_F(t)$ and $\eta_B(t)$ are the fields that lie on the forward ($\eta_F$)
  and the backward ($\eta_B$) sides of the Keldysh contour
	(for more details see  \ref{sec:app_nlo} ).

 It turns out convenient to rotate the fields $\eta_F(t)$ and $\eta_B(t)$ to so-called "classical" \mbox{$\phi_c = \frac{1}{2}(\eta_F +\eta_B)$} and "quantum"  \mbox{$\phi_q = \eta_F-\eta_B$} components
:
 \begin{gather}\label{Tmm_1}
  \langle T^{\mu}_{\mu}\rangle_{t_1} = \int d\xi\ D[\xi_1,\rho_0,\xi_2]
  \int \limits_{\varphi_c(t_0) = \frac{\xi_1+\xi_2}{2}}^{\varphi_c(\infty) = \xi}
  \mathcal{D}\varphi_c 	 \\ \times
 	\int \limits_{\varphi_q(t_0) =\xi_1-\xi_2}^{\varphi_q(\infty) = 0}
  \mathcal{D}\varphi_q\  e^{i S_K[\varphi_c,\varphi_q]}
  \Big(- \dot\varphi_c^2 + g^2 \varphi_c^4\Big). \nonumber
 \end{gather}

The Keldysh action for the  theory (\ref{stat_lagr}) reads
\begin{multline}
	S_K[\phi_c,\phi_q] =
  V_3\int\limits_{t_0}^{\infty}dt \Big(\dot\phi_c\dot\phi_q
  -g^2\phi_c^3\phi_q \\
	-\frac{g^2}{4}\phi_c\phi_q^3 + J\phi_q  \Big).
\end{multline}
 The variation over $\phi_q$ is
 \begin{gather}
	 \frac{\delta S_K}{\delta \varphi_q}\Big|_{J=0} =  - V_3 \big(\ddot \varphi_c
	 + g^2 \varphi^3_c + \frac{3}{4}g^2 \varphi_c\varphi_q^2 \big)
 \end{gather}
and, therefore, the equation (\ref{Tmm_1}) for the trace of the energy-momentum tensor can be rewritten in the following form:
\begin{multline}\label{Tmm_2}
	\langle T^{\mu}_{\mu}\rangle_{t_1} = \int d\xi\ D[\xi_1,\rho_0,\xi_2]
	\int \mathcal{D}\varphi_c 	\mathcal{D}\varphi_q
	\Big( - \frac{\varphi_c}{V_3}
	\frac{\delta S_K}{\delta \varphi_q}\\
	 -	\frac{3}{4}g^2 \varphi_c^2\varphi_q^2
	- \dot\varphi_c^2  -\varphi_c \ddot\varphi_c
	 \Big)e^{i S_K[\varphi_c,\varphi_q]}.
\end{multline}

The first term of in (\ref{Tmm_2}) can be shown to vanish by integrating by parts and neglecting the surface term.  The second term vanishes because the
considered observable depends only on one time variable, and,
therefore,
\begin{multline}
\int \mathcal{D}\eta_F\mathcal{D}\eta_B\ \eta_F(t_1)e^{i S_K[\eta_F,\eta_B]}= \\
\int \mathcal{D}\eta_F\mathcal{D}\eta_B\ \eta_B(t_1)e^{i S_K[\eta_F,\eta_B]} .
\end{multline}
we see that all terms with $\varphi_q^n\equiv (\eta_F-\eta_B)^n$ disappear. The last two terms can be expressed through the total time derivative, so that the final expression for $\langle T^{\mu}_{\mu}\rangle_{t_1}$ takes the form
\begin{multline}\label{Tmm_full}
	\langle T^{\mu}_{\mu}\rangle_{t_1} =   -\frac{1}{2}
	\int d\xi\ D [\xi_1,\rho_0,\xi_2]  \\
	\times\int \mathcal{D}\varphi_c
	\mathcal{D}\varphi_q\  e^{i S_K[\varphi_c,\varphi_q]}\
	 \partial^2_{t_1} \varphi_c^2(t_1) \\
	 \equiv
	 -\frac{1}{2}\partial^2_{t_1}\langle \hat \varphi^2(t) \rangle_{t_1}.
\end{multline}

Let us stress that the above expression \eqref{Tmm_full}  is exact.
It describes full quantum evolution of the trace of energy-momentum tensor
 $T_{\mu}^{\mu}$. Intuitively at large enough time, when the field
 equilibrates to some constant value,
  the trace of energy-momentum tensor should vanish due to time derivative
	. In the static
	geometry case this will indeed be shown below by analytical calculation
	of
	$\langle T^{\mu}_{\mu}\rangle$ at the leading and next-to-leading
	order
	in quantum corrections to the classical approximation.
	 As shown in detail  in the
	  \ref{sec:app_nlo}, the expression for $\langle T^{\mu}_{\mu}\rangle$
		 in the leading and next-to-leading approximation
		 of the semiclassical expansion reads
 \begin{multline}\label{Tmm_sem_st}
	\langle T^{\mu}_{\mu}\rangle_{t_1} =   -\frac{1}{2} \partial^2_{t_1}
	\Bigg< \phi_{cl}^2(t_1)  \\
	+ \frac{g^2}{4 V_3^2} \int\limits_{t_0}^{t_1}dt_2 \phi_{cl}(t_2)
\frac{\delta^3\phi^2_{cl}(t_1)}{\delta J(t_2)^3}\Bigg|_{J=0}
	 \Bigg>_{i.c.},
\end{multline}
where  $\phi_{cl}(t)$ is the solution (\ref{class_sol}) of the EoM, and brackets $\langle\rangle_{i.c.}$ denote integration over  initial conditions with the weight given by the  Wigner function $f_W(\alpha,p,t_0)$
\begin{gather}\label{ic_def}
 \langle F(t) \rangle_{i.c.} = \int d\alpha dp f_W(\alpha,p,t_0) F(t),\nonumber\\
 f_W(t_0,\alpha,p) = \int d\beta \langle\alpha+\frac{\beta}{2}
 |\hat\rho(t_0)|\alpha-\frac{\beta}{2} \rangle e^{i V_3 \beta  p},\\
  \alpha = \phi_{cl}(t_0), \nonumber\\
 p=\partial_t \phi_{cl}(t_0).
\end{gather}
 Let us note that the first term in (\ref{Tmm_sem_st}) corresponding to the leading order (LO) quantum correction matches with the Classical Statistical Approximation.

 Let us first work out an expression for the LO term in \eqref{Tmm_sem_st}. Due to periodicity of the classical solution (\ref{class_sol})
 \begin{gather}
  \phi_{cl}(t) = \phi_m \sum\limits_{k=-\infty}^{\infty} u_k
  e^{\frac{2\pi  i k}{T_{cl}}(g \phi_m t +C)},\\
  u_k = \frac{1}{T_{cl}}\int\limits_{0}^{T_{cl}} cn\Big(\frac{1}{2},t\Big)
  e ^{-\frac{2\pi i k t}{T_{cl}}}\nonumber
 \end{gather}
it is possible to calculate the LO term in  (\ref{Tmm_sem_st}) analytically with the Gaussian Wigner function ansatz
 \begin{gather}\label{wigner}
  f_W(\alpha,p,t_0)=\frac{1}{\alpha_0 p_0 \pi}e^{-\frac{(\alpha-A)^2}{\alpha_0^2}}
  e^{-\frac{p^2}{p^2_0}}.
 \end{gather}
Note that the amplitude $\phi_m $ and phase $C$ of the classical trajectory are functions of the initial conditions
\begin{gather}
 \alpha = \phi_m cn\Big(\frac{1}{2},C\Big),\quad
p = -g\phi_m^2 sn \Big(\frac{1}{2},t\Big)\cdot dn\Big(\frac{1}{2},t\Big),
\label{icond}
\end{gather}
where $sn(k^2,x)$ and $dn(k^2,x) $ are the Jacobi elliptic functions.
With help of relations \eqref{icond} we can replace integration over initial conditions with that over possible amplitudes and phases of the trajectory
$\int d\alpha dp \to \int d\phi_m dC $ and perform the integration in the saddle point approximation ($\phi_m = A,\ C=0$). The resulting expression for the LO contribution in  (\ref{Tmm_sem_st})  then reads
\begin{gather}\label{ic_lo}
 \langle T_{\mu}^{\mu} \rangle_{t_1}^{LO}\equiv
 -\frac{1}{2}\partial_{t_1}^2 \langle \phi_{cl}(t_1)^2\rangle_{i.c.}=\\
  -\frac{1}{2}\partial_{t_1}^2 \Big( A^2 \sum\limits_{k=-\infty}^{\infty}
   I(k) e^{-\frac{\pi^2 p_0^2 k^2}{g^2 A^4 T_{cl}^2}}
   e^{-\frac{\pi^2 \alpha_0^2 g^2 k^2  t_1^2}{ T_{cl}^2}}
   e^{\frac{i 2\pi  A g k t_1}{T_{cl}}} \Big)\nonumber,\\
   I(k) = \sum\limits_{n=-\infty}^{\infty}u_n u_{k-n}
   = \frac{1}{T_{cl}}\int\limits_0^{T_{cl}}cn^2\Big(\frac{1}{2},t\Big)
   e^{-\frac{2\pi i k}{T_{cl}}} dt\nonumber .
\end{gather}
 The parameter $A$ of the Wigner distribution (\ref{wigner}) is a measure of the field intensity. As  shown in the previous papers \cite{LR1,LR2},  the large $A$ limit is directly related  to the  validity of the CSA.  In what follows we show that quantum corrections to the CSA (or next-to-leading order of the semiclassical decomposition) scale as $A^{-n}$ and, therefore, vanish in the large $A$ limit.

After averaging over initial conditions equation \eqref{ic_lo} contains three types of  exponents: constant in time,  oscillating and decaying as $e^{-t^2}$. Obviously, in the large time limit
$t\to\infty$ the LO part does vanish. The only dangerous term in the sum is the one with $k=0$. However, as this term is time-independent, it vanishes after differentiation over time in Eq.~(\ref{ic_lo}).

Let us now consider the NLO term in Eq.~(\ref{Tmm_sem_st})
\begin{multline}\label{nlo_st_1}
 \langle T_{\mu}^{\mu} \rangle_{t_1}^{NLO} =
 -\frac{1}{2}\partial_{t_1}^2 \Bigg\langle
\frac{g^2}{2V_3^2} \int\limits_{t_0}^{t_1} dt_2 \phi_{cl}(t_2) \\
\times
\Big( \phi_{cl}(t_1)\Phi_3(t_1,t_2)
+3\Phi_1(t_1,t_2)\Phi_2(t_1,t_2)
 \Big)\Bigg\rangle_{i.c.},
\end{multline}
where $\Phi_n(t_1,t_2)$ are variations of the classical EoM over
the auxliary source $J$. They can be shown to satisfy the
following differential equations
(see  \ref{sec:app_nlo}):
\begin{gather}\label{variationss_st}
\Phi_n(t_1,t_2) = \frac{\delta^n \phi_{cl}(t_1)}{\delta J^n(t_2)}, \\
 L_{t_1}\Phi_1(t_1,t_2) = \delta(t_1-t_2), \nonumber\\
 L_{t_1}\Phi_2(t_1,t_2) = -6 g^2 \phi_{cl}(t_1)\Phi_1^2(t_1,t_2),\nonumber\\
 L_{t_1}\Phi_3(t_1,t_2) = -6 g^2 \Phi_1^3(t_1,t_2)\nonumber \\
 - 18 g^2 \phi_{cl}(t_1)\Phi_1(t_1,t_2)\Phi_2(t_1,t_2),\nonumber\\
 L_{t_1} = \partial^2_{t_1} + 3 g^2 \phi_{cl}^2(t_1). \nonumber
\end{gather}

It turns out convenient to define a dimensionless variable  $z = g \phi_m t + C $ and
  dimensionless variations $f_n$ as
 \begin{multline}\label{var}
  \phi_{cl}(t)\to\phi_m f_0(z),\\
  \Phi_n(t_1,t_2)\to g^{-n} \phi_m^{1-3n} f_n(z_1,z_2), \qquad n=1,2,3.
\end{multline}
The equation \eqref{nlo_st_1} takes the form
\begin{multline}\label{nlo_st_2}
 \langle T_{\mu}^{\mu} \rangle_{t_1}^{NLO} =
 -\frac{1}{2}\partial_{t_1}^2 \Bigg\langle
\frac{1}{2V_3^2 g^2 \phi_m^4} \int\limits_{z_0}^{z_1} dz_2 f_0(z_2) \\
\times
\Big( f_0(z_1)f_3(z_1,z_2)
+3f_1(z_1,z_2)f_2(z_1,z_2)
 \Big)\Bigg\rangle_{i.c.}.
\end{multline}
Using the integral representation of equations (\ref{variationss_st})
 one can show that in the limit $z_1-z_2\to\infty$ the functions
 $f_n$ scale as
\begin{gather}
 f_n(z_1,z_2)\vert_{z_1-z_2\to\infty} \approx (z_1-z_2)^n.
\end{gather}
Hence, the dimensionless integral in \eqref{nlo_st_2} can be rewritten as
\begin{multline}\label{F_st}
 \int\limits_{z_0}^{z_1} dz_2 F_{nlo}(z_1,z_2)
 \equiv \int\limits_{z_0}^{z_1} dz_2 f_0(z_2)
\Big( f_0(z_1)f_3(z_1,z_2)\\
+3f_1(z_1,z_2)f_2(z_1,z_2)\Big) =
\sum\limits_{n=0}^3 \psi_n(z_1) z_1^n,
\end{multline}
where $\psi_n(z_1)$ are periodic functions (with period equal to $T_{cl}$) which can be found  numerically.

We can use Fourier transform of these periodic functions
\begin{gather}
	\psi_n(z)  =\sum\limits_{k=-\infty}^{\infty} \psi_n^{(k)}
	e^{i k z \frac{2\pi}{T_{cl}}}
\end{gather}
to perform integration over initial condition using the same
 method as for the LO calculations (\ref{ic_lo}).
  The final expression for the trace of the energy-momentum tensor
	including LO and NLO contributions in quantum corrections of the
	semiclassical expansion does then read
\begin{multline}\label{ic_nlo}
	\langle T_{\mu}^{\mu} \rangle_{t_1}^{LO+NLO} =
   -\frac{A^2}{2}\partial_{t_1}^2   \times \\
  \sum\limits_{k=-\infty}^{\infty} \Big(  I(k) +
	\frac{1}{2 V_3^2 g^2 A^6}  \sum\limits_{n=0}^{3}
	 \psi_n^{(k)} (g A t_1)^n
	\Big)\times\\
	 e^{-\frac{\pi^2 p_0^2 k^2}{g^2 A^4 T_{cl}^2}}
    e^{-\frac{\pi^2 \alpha_0^2 g^2 k^2  t_1^2}{ T_{cl}^2}}
		   e^{\frac{i 2\pi  A g k t_1}{T_{cl}}}.
\end{multline}
From equation \eqref{ic_nlo} we see that at large times the trace of the energy-momentum tensor does indeed vanish. The only subtlety is again related to the  zero  Fourier components
of the periodic  functions  $\psi_n^{(0)}$.  However, it is easy to restore these functions numerically using evaluated value of the integral (\ref{F_st}) and the Vandermonde matrix. This calculation shows that
all the zeroth Fourier components vanish $ \psi_n^{(0)} = 0$ and, therefore, the trace of the energy-momentum tensor does indeed relax to zero at large enough observation time $t_1$.

This fact very important for working out physical interpretation of the studied evolution of nonequilibrium quantum field. Vanishing of the trace of energy-momentum tensor means that there establishes a well defined relation between energy density and pressure, i.e. the equation of state thus making it possible to work out a hydrodynamics description of the dynamics under consideration.

Let us note that from  the expression (\ref{ic_nlo}) we see that significant contributions form the NLO terms correspond to the limit of small $A$. Therefore for the CSA approximation to be valid  we need to choose the
Wigner distributions with  large initial amplitudes
$\phi_m(\alpha,p)$ and fast decaying tales \cite{LR1,LR2}.

\section{\label{sec:expanding}expanding geometry}

Let us now turn to the analysis of evolution of energy-momentum tensor
in the case of the geometry expanding in the longitudinal direction \cite{initial3}.  The natural coordinates  describing a system undergoing longitudinal expansion along the z axis are
\begin{gather}
 \tau^2 = t^2 - z^2,\\
 \eta = \frac{1}{2}\log\frac{t+z}{t-z},\nonumber\\
 {\bf x_{\perp}} = {\bf x_{\perp}}.\nonumber
\end{gather}
As before, we consider the spatially homogeneous case,
 $\partial_{\eta}\varphi = 0$ and $\partial_{\perp}\varphi = 0$.
The action for the case of expanding geometry reads
\begin{gather}
 S = V_2 \int d\tau\ \tau \Big(
  \frac{1}{2} \dot \varphi^2 - \frac{g^2}{4}\varphi^4+ J \varphi\Big), \nonumber\\
 V_2 = \int d^2 x_{\perp}d\eta,
\end{gather}
where $\dot\varphi \equiv \frac{\partial \varphi}{\partial \tau}$.
The classical trajectories are given by the solutions of the following EoM
 \begin{gather}\label{class_exp}
 \ddot \phi_{cl.e} + \frac{1}{\tau}\dot \phi_{cl.e} + g^2 \phi^3_{cl.e} = 0
\end{gather}
equipped with certain initial conditions. The subscript "e" stands for "expanding" and refers to the values related to the expanding coordinate system.
The EoM \eqref{class_exp} for the expanding case does not allow the analytical solution. However, using the substitution
\begin{gather}
	y=\tau^{\frac{2}{3}},\\
	\phi_{cl.e}(\tau) = \tau^{-\frac{1}{3}}\xi(\tau^{\frac{2}{3}}),\nonumber
\end{gather}
which effectively takes into account the expansion rate, one can see that the EoM \eqref{class_exp} in new variables
\begin{gather}\label{EoM_expanding_y}
\ddot \xi(y)+\frac{1}{4y^2} \xi(y) +\frac{9}{4} g^2 \xi(y)^3 = 0
\end{gather}
does at large times take the form of the one for the static geometry (\ref{EoM}) and thus possess in this limit an asymptotic analytical solution of the form
\begin{gather}
	\xi(y) \to \xi_m cn\Big(\frac{1}{2}, \bar g \xi_m y + C_{\xi}\Big),
\end{gather}
where $\bar g = \frac{3}{2}g$ and $\xi_m,\ C_{\xi}$ are correspondingly the amplitude and the phase characterizing the asymptotic periodic trajectory. Let us denote by $\tilde y$ the "time" where this periodic regime sets in. As one can see from eq.(\ref{EoM_expanding_y}) this "periodization" scale $\tilde y$ decreases with increasing coupling constant and/or field amplitude.  Let us note that these conditions are similar to those controlling the validity of the CSA. The corresponding asymptotic solution of the classical EoM (\ref{class_exp}) for $\tau > \tilde\tau = \tilde y^{\frac{3}{2}}$ then reads
\begin{gather}\label{tilde_phi_e}
	\tilde\phi_{cl.e}(\tau) = \xi_m \tau^{-\frac{1}{3}}
	cn\Big(\frac{1}{2}, \bar g \xi_m \tau^{\frac{2}{3}} + C_{\xi} \Big).
\end{gather}
It is important to note that  presence of the small initial time interval $0<\tau<\tilde \tau$ in which the solution is not periodic precludes us from establishing analytical relation between the initial condition ($\alpha = \phi_{cl.e}(t_0)$, $p = \dot\phi_{cl.e}(t_0) $) and the parameters of the  trajectory ($\xi_m$, $C_{\xi}$).

At tree level the expression for the trace of the energy momentum tensor reads
\footnote{Note that $\phi_{cl.e}$ is an exact solution of eq. (\ref{class_exp}) whereas $\tilde\phi_{cl.e}$ of (\ref{tilde_phi_e}) corresponds to the approximate periodic-like solution.}
\begin{gather}\label{Tmm_class_exp}
		\big(T^{\mu}_{\mu}\big)_{cl}^e = -\dot\phi^2_{cl.e} + g^2 \phi^4_{cl.e}
\end{gather}

The quantum evolution is described in  the same way as in the static case eq.(\ref{Tmm_full})
\begin{multline}\label{Tmm_full_e}
	\langle T^{\mu}_{\mu}\rangle_{\tau_1} =
	-\frac{1}{2}
 \int d\xi\ [\xi_1,\rho_0,\xi_2]  \\
 \times\int \mathcal{D}\varphi_c
 \mathcal{D}\varphi_q\  e^{i S_K^e[\varphi_c,\varphi_q]}\
\Big(\partial^2_{\tau_1}+\frac{1}{\tau_1}\partial_{\tau_1} \Big)
	\varphi_c^2(\tau_1) \\
	\equiv
	 -\frac{1}{2}\Big(\partial^2_{\tau_1}+\frac{1}{\tau_1}\partial_{\tau_1} \Big)
	 \langle  \phi^2_{cl.e}(\tau_1) \rangle_{\tau_1},
\end{multline}
with the following Keldysh action in the expanding coordinates
\begin{multline}
	S_K^e[\phi_c,\phi_q] =
  V_2\int\limits_{\tau_0}^{\infty}d\tau\ \tau\  \Big(\dot\phi_c\dot\phi_q
  -g^2\phi_c^3\phi_q \\
	-\frac{g^2}{4}\phi_c\phi_q^3 + J\phi_q  \Big).
\end{multline}

The asymptotic large $\tau_1$ behavior of the quantity $ \langle  \phi^2_{cl.e}(\tau_1) \rangle_{\tau_1}$  in eq.(\ref{Tmm_full_e})  is
$ \langle  \phi^2_{cl.e}(\tau_1) \rangle_{\tau_1} \sim \tau_1^{-\frac{2}{3}}$ and,
 therefore, $T_{\mu}^{\mu}=0$ for $\tau\to\infty$ to all orders in the semiclassical expansion.
  However, if we take into account the expansion rate, we can find an intermediate quasistationary regime with a nontrivial equation of state.  To describe this regime it turns out convenient to rescale the expression for the averaged trace of the energy-momentum tensor in (\ref{Tmm_full_e}) by dividing it on the LO energy
 \begin{gather}
 \varepsilon_{LO}^e = \Big\langle\frac{1}{2}\dot\phi_{cl.e}^2
 +\frac{g^2}{4}\phi_{cl.e}^4\Big\rangle_{i.c.}^e
 \sim \tau^{-\frac{4}{3}},
 \end{gather}
thus effectively removing the influence of the expansion rate, see eq. (\ref{Tmm_sem_e}) below.

The  averaging over initial conditions in the expanding  case is described by
\begin{gather}\label{ic_def_e}
 \langle F(t) \rangle_{i.c.}^e =
 \int d\alpha dp f_W^e(\alpha,p,t_0) F(t),\\
 f_W^e(\tau_0,\alpha,p) = \int d\beta \langle\alpha+\frac{\beta}{2}
 |\hat\rho(\tau_0)|\alpha-\frac{\beta}{2} \rangle
 e^{i V_2\tau_0 \beta  p},\nonumber\\
  \alpha = \phi_{cl.e}(\tau_0), \nonumber\\
 p=\partial_\tau \phi_{cl.e}(\tau_0)\nonumber.
\end{gather}
The semiclassical decomposition for the expanding case reads
\begin{multline}\label{Tmm_sem}
	\langle T^{\mu}_{\mu}\rangle_{t_1} =   -\frac{1}{2}
	\Big(\partial^2_{\tau_1}+\frac{1}{\tau_1}\partial_{\tau_1} \Big)
	\Bigg< \phi_{cl.e}^2(\tau_1)  \\
	+ \frac{g^2}{4 V_2^2} \int\limits_{\tau_0}^{\tau_1}
	\frac{d\tau_2}{\tau_2^2}\phi_{cl.e}(\tau_2)
\frac{\delta^3\phi^2_{cl.e}(\tau_1)}{\delta J(\tau_2)^3}\Bigg|_{J=0}
	 \Bigg>_{i.c.}^e.
\end{multline}
The equations for variations $\Phi_n^e(\tau_1,\tau_2)$ are similar to the ones in the static case (\ref{variationss_st}), albeit with a different differential operator
\begin{gather}\label{variationss_e}
 L_{t}\to L_{\tau}=\partial_{\tau}^2 +\frac{1}{\tau}\partial_{\tau} + 3 g^2 \phi_{cl.e}^2 .
\end{gather}
We write expression for the trace
of the energy-momentum tensor at the NLO accuracy
 as a sum of two contributions - the initial aperiodic,
 corresponding to the time interval [$\tau_0,\tilde \tau$], and asymptotic periodic
  corresponding to the interval [$\tilde \tau,\tau_1$].

Let us make the following substitutions:\\
-  to take into account the effects of expansion
\begin{gather}
 	y=\tau^{\frac{2}{3}},\\
	\phi_{cl.e}(\tau) = \tau^{-\frac{1}{3}}\xi(\tau^{\frac{2}{3}}),\nonumber\\
	\Phi_n^e(\tau_1,\tau_2)=
\left( \frac{3}{2}\right)^n \tau_1^{-\frac{1}{3}}
\tau_2^{\frac{2n}{3}} \rho_n
\left(\tau_1^{\frac{2}{3}},\tau_2^{\frac{2}{3}}\right),\quad n=1,2,3\nonumber.
\end{gather}
- to make relevant quantities dimensionless
\begin{gather}\label{var_exp}
 z_e = \bar g \xi_m y +C_{\xi},\\
 \xi(y) = \xi_m f_0^e(z^e),\nonumber\\
 \rho_n(y_1,y_2) = \bar g^{-n}\xi_m^{1-2n} f_n^e(z_1^e,z_2^e)\nonumber.
\end{gather}
The resulting rescaled expression for the trace of the energy-momentum tensor then reads
\begin{gather}
		\frac{\langle T^{\mu}_{\mu}\rangle_{\tau_1}^{LO+NLO}}{ \varepsilon_{LO}^e} =
		-\frac{1}{2 \varepsilon_{LO}^e(\tau_1)}
		\Big( \partial_{\tau_1}^2 +
		\frac{1}{\tau_1}\partial_{\tau_1}\Big)\Bigg\langle\phi_{cl.e}^2(\tau_1)
		+ \nonumber\\
		\frac{\tau_1^{-\frac{2}{3}}}{2 V^2_2 g^2 \xi_m^4} \Big(
		\int\limits_{ z_0^e}^{ \tilde z}d z_2^e F_{nlo}^e(z_1^e,z_2^e)+
		\int\limits_{ \tilde z}^{  z_1^e}d z_2^e F_{nlo}^e(z_1^e,z_2^e)
		\Big)\Bigg\rangle_{i.c.}^e.
		\label{Tmm_sem_e}
\end{gather}

In these new notations the "periodization" scale $\tilde y$ turns into $\tilde z = \bar g \xi_m \tilde y +C_{\xi} $.
After  time  $\tilde z$ the dimensionless variations (\ref{var_exp}) become periodic (of the form of (\ref{var}))
\begin{gather}
 f_k^e(z_1^e,z_2^e) =  f_k(z_1^e,z_2^e) , \quad z^e_2 >\tilde z.
\end{gather}
There follows that the dimensionless integral over the asymptotic periodic-like interval
($\tilde z, z_1^e$)
\begin{gather}
 \int\limits_{\tilde z}^{z_1^e}dz^e_2 F_{nlo}^e(z_1^e,z_2^e) \equiv
 \int\limits_{\tilde z}^{z_1^e}dz^e_2 \Big( f_0(z_1^e)f_3(z_1^e,z_2^e)\\
+3f_1(z_1^e,z_2^e)f_2(z_1^e,z_2^e)\Big)
\end{gather}
becomes similar to its static analogue (\ref{F_st}).
Therefore, one can use the same arguments as in the previous section
in order to show that after averaging over the initial conditions
the last  term in  eq.(\ref{Tmm_sem_e}) vanishes at large times.

The LO contribution (the first term in eq. \eqref{Tmm_sem_e} )
also vanishes after averaging due to periodicity
at the large-times.  As it is shown in \cite{initial1} this statement
about the LO contribution can be proved with the other arguments as well.

 Therefore, the large-time behaviour of the trace
 of the energy-momentum tensor is governed by the second term of
   eq.\eqref{Tmm_sem_e} which include integration over
	  the initial time interval [$z_0^e,\tilde z$]

 \begin{gather}
		\frac{\langle T^{\mu}_{\mu}\rangle_{\tau_1}^{LO+NLO}}{ \varepsilon_{LO}^e}
		 \underset{\tau\gg \tilde \tau}{=}
		-\frac{1}{2 \varepsilon_{LO}^e(\tau_1)}
		\Big( \partial_{\tau_1}^2 +
		\frac{1}{\tau_1}\partial_{\tau_1}\Big)\times\nonumber\\
		\Bigg\langle
		\frac{\tau_1^{-\frac{2}{3}}}{2 V^2_2 g^2 \xi_m^4}
		\int\limits_{ z_0^e}^{ \tilde z}d z_2^e F_{nlo}^e(z_1^e,z_2^e)\Bigg\rangle_{i.c.}^e.
		\label{Tmm_sem_e_2}
\end{gather}
The above expression manifest differences between  static and longitudinally expanding theories.
This term breaks scale invariance (see discussion  in section \ref{sec:results})
that can lead to a nonzero contribution to the trace of the energy-momentum tensor.
It is not so easy to perform the
 integration over initial condition analytically in eq. \eqref{Tmm_sem_e_2}. However, we expect
 this term to be suppressed by the intensity of the initial field (parameter $A$ in eq. \eqref{ic_nlo});
 to have decaying, constant and growing with time parts.

 \section{\label{sec:numerical} Numerical results }

 In this section we present the results of the numerical calculations for the trace
of the energy-momentum tensor in the expanding background.
  These calculations are made with formula

	\begin{multline}
	  \langle T^{\mu}_{\mu}\rangle_{\tau_1}^{LO+NLO} =
		\Bigg\langle T^{\mu}_{\mu}(\tau_1)_{cl.e} \\
	+
	  \frac{g^2}{4 V_2^2}\int\limits_{\tau_0}^{\tau_1}\frac{d\tau_2}{\tau_2^2}
	  \phi_{cl.e}(\tau_2)\frac{\delta^3\ T^{\mu}_{\mu}(\tau_1)_{cl.e} }{\delta J^3(\tau_2)}
		 \Bigg\rangle_{i.c.}^e \label{exp_mean}
	\end{multline}
	followed  from the definition of the NLO corrections
	without additional assumptions (see \ref{sec:app_nlo}).
	The classical solutions $\phi_{cl.e}$ and the trace $T^{\mu}_{\mu}(\tau_1)_{cl.e}$
	are given by formulae \eqref{class_exp} and \eqref{Tmm_class_exp}
	respectively. The variation over additional source reads
	\begin{multline}
		\frac{\delta^3\ T^{\mu}_{\mu}(\tau_1)_{cl.e} }{\delta J^3(\tau_2)}=
		4g^2\big[ 6 \phi_{cl.e}(\tau_1)\big(\Phi_1^e(\tau_1,\tau_2) \big)^3\\
	+	9 \phi_{cl.e}^2(\tau_1) \Phi^e_1(\tau_1,\tau_2)\Phi^e_2(\tau_1,\tau_2)
		+ \phi^3_{cl.e}(\tau_1) \Phi^e_3(\tau_1,\tau_2)\big]\\
		- 2 \big(\dot\phi_{cl.e}(\tau_1)\dot\Phi^e_3(\tau_1,\tau_2)
		+ 3\dot\Phi^e_1(\tau_1,\tau_2)\dot\Phi^e_2(\tau_1,\tau_2) \big) ,
	\end{multline}
where functions $\Phi_n(\tau_1,\tau_2)$ are the solutions of the differential
equations \eqref{variationss_e} and \eqref{variationss_st}, the dot means
derivative with respect to $\tau_1$.
Averaging over the ensemble of the initial condition is done with
the Gaussian Wigner function
(\ref{wigner}).

   Simulations are performed at different values of the parameter $A$
  of the initial distribution \eqref{wigner}. This parameter defines the intensity of
the initial field for the described homogeneous model or, in other words, the applicability of
the semiclassical decomposition.

The figures are organised as follows: the top panel shows the evolution
of the trace $T_{\mu}^{\mu}$ as a function of time $\tau_1$,
the bottom one shows the ratio of the NLO energy density to LO energy
density
which demonstrate the applicability of the  semiclassical decomposition.

\begin{figure}
      \includegraphics[width=0.48\textwidth]{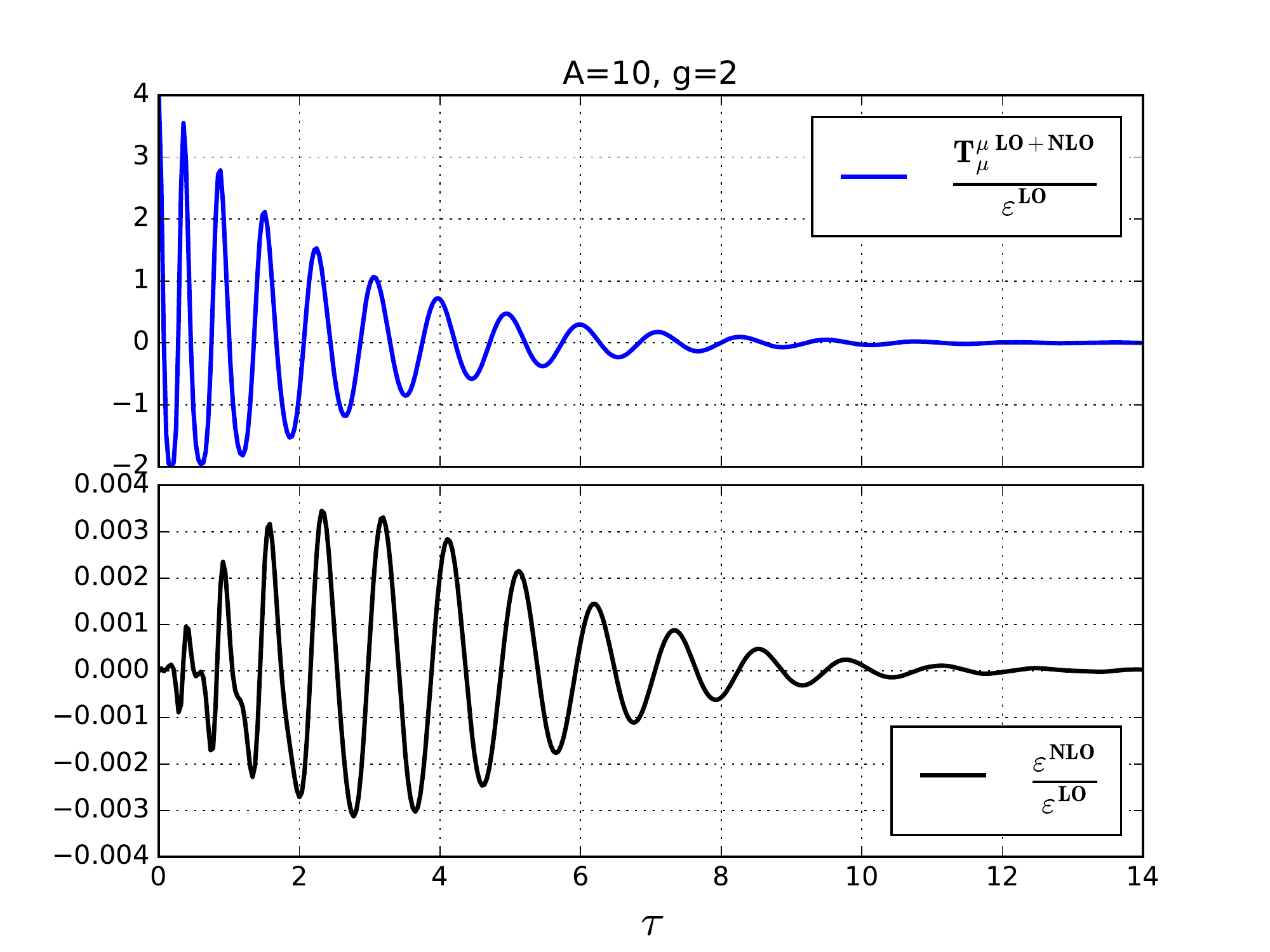}
    \caption{Parameters of the Wigner function \eqref{wigner}:
		$A$=10, $\alpha_0=p_0=1$, $g$=2.
		Top panel: The trace of the energy-momentum
		tensor \eqref{exp_mean} normalised by LO energy density; bottom panel:
		the ration of the NLO energy density and the LO energy density}
		\label{pA10}
        \end{figure}

 Fig. \ref{pA10}   shows the case in which CSA works extremely well. The parameter $A$ is large enough to
 neglect NLO term contribution at large times.  Hence, the trace
 of the energy-momentum tensor averages to a very small constant.

\begin{figure}
  \includegraphics[width=0.48\textwidth]{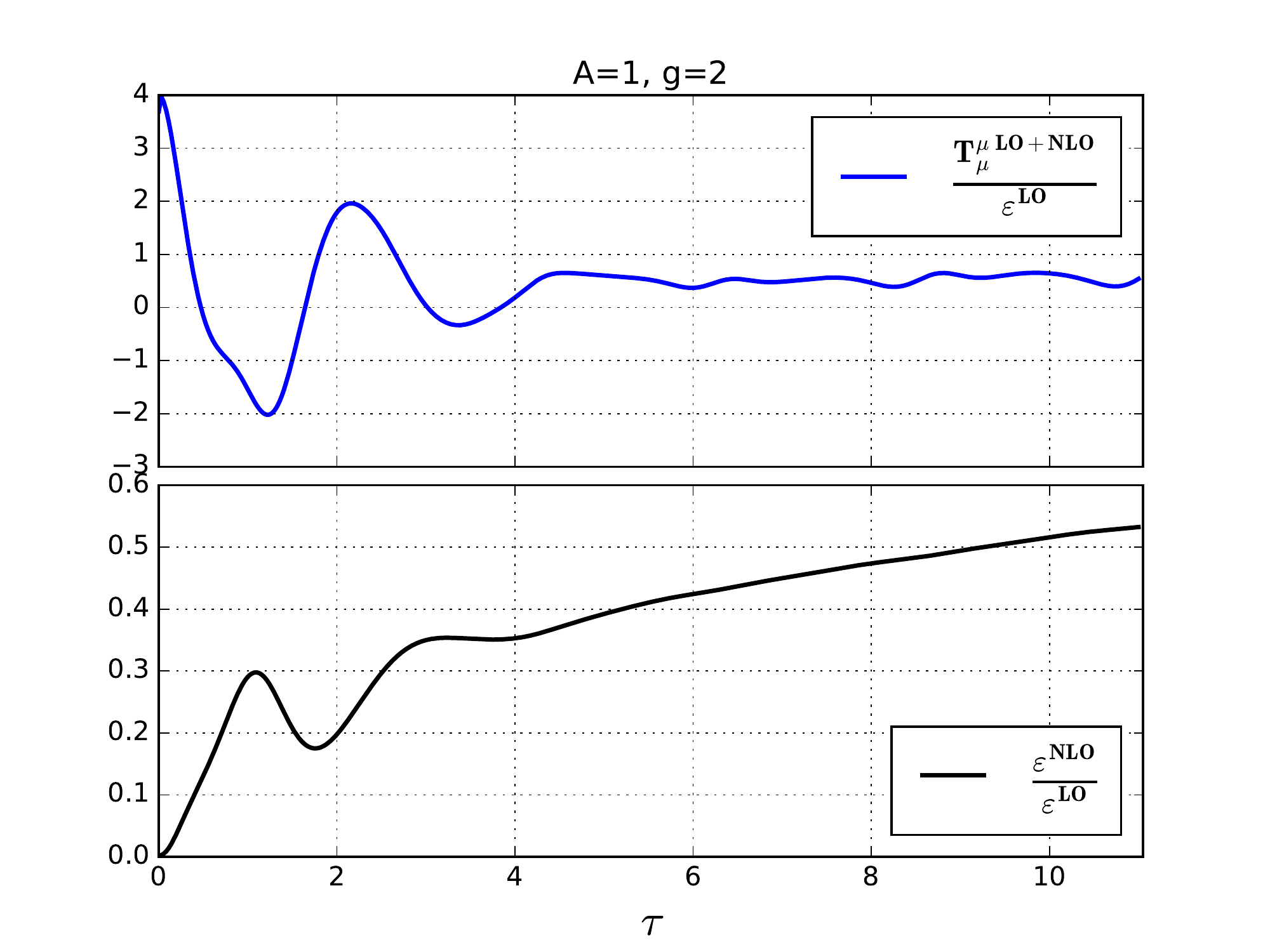}
    \caption{The same as for Fig.\ref{pA10} but with $A$=1}
		\label{pA1}
      \end{figure}

 Fig. \ref{pA1} demonstrates the evolution of the trace of
 energy-momentum tensor for the "intermediate" range of the
 initial parameters , where semiclassical decomposition is still
 adequate, but the NLO corrections are already important.
 One can observe a regime  with $T_{\mu}^{\mu}\propto\varepsilon$ indicating formation of the equation of state of the form of  $\varepsilon = c p$ with $c> 3$. This result demonstrates a direct nontrivial effect of the NLO quantum corrections.

\begin{figure}
        \includegraphics[width=0.48\textwidth]{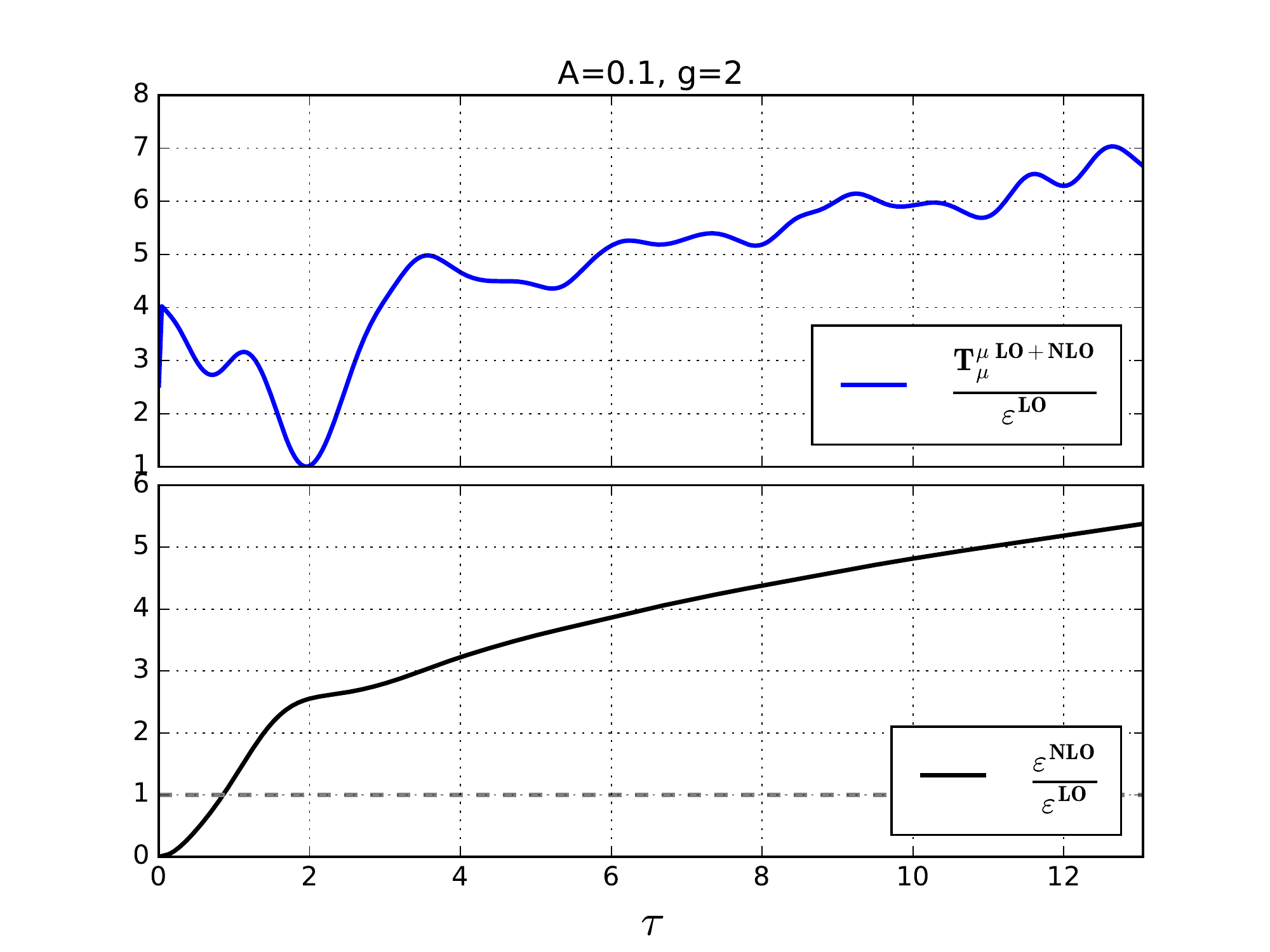}
    \caption{The same as for Fig.\ref{pA10} but with $A$=0.1}
		  \label{pA01}
  \end{figure}

Fig. \ref{pA01} shows the result for the deeply quantum case in which the CSA approximation does not hold.

  \section{\label{sec:results} Conclusions}

We calculated the quantum corrections to the trace of the energy-momentum tensor for the homogeneous $\varphi^4$ scalar field in two cases.

In the first case, the field placed inside a static box,
 we demonstrated that the NLO quantum corrections give contributions that vanish
 at large enough times.  As one can see from lagrangian \eqref{4scalar} the system posses the scale invariance which is broken by the initial
conditions. However, during the time evolution the system forgets about initial state due to the self-interaction. That is why at the large times the scale invariance restores, $T_{\mu}^{\mu}=0$ and the quantum correction are negligible.

In the second case, the longitudinally expanding field,
we claimed that the quantum correction might change the meanvalue of
the trace of the energy-momentum tensor.  In other words,
the intermediate quasistationary regime characterised by the equation
of state of the form different from the relativistic one $\varepsilon=3p$
is realised. This regime exists in the expanding geometry because
the system transforms from the classical to the quantum
one during the expansion. This phenomena occurs for the certain
range of the parameters characterising the distribution of the
initial conditions.

 Note that the nonzero value of the trace of the
 energy-momentum tensor seen in Fig.~(\ref{pA1})
 does not contradict with the scale invariance mentioned above.
 The scale invariance results in the requirement of
 $\langle T^{\mu}_{\mu} \rangle =0$ in equilibrium.
 However, the new regime we observed is an intermediate
 nonequilibrium one. Asymptotically, the system
 evolves to the equilibrium state with\\
 $\varepsilon =0,\quad p=0$
 due to expansion, hence scale invariance is restored.

 In this work we describe the oversimplified scalar system. However, the phenomena we observed might be valuable for the
 description of the ultrarelativistic heavy-ion collisions as well.   We suggest that similar quasistationary state can be formed
 due to  nonequilibrium conditions, which are present in the matter created in such collisions.

\appendix
\section{\label{sec:app_nlo} Quantum corrections to the CSA: scalar field theory}

In this Appendix we describe a general formalism for calculation of
quantum corrections to the Classical Statistical Approximation.
For simplicity we consider the case of the scalar field, however,
the idea can be extended to gauge fields as well.
The main observation is that in the Keldysh-Schwinger formalism
the CSA represents the Leading Order term of the semiclassical decomposition
thus providing a basis for the systematic expansion.

Out of equilibrium an expectation value of observable $F(\hat\varphi)$
at the moment $t_1$
can be calculated  as a trace with density matrix as
\begin{multline}
  \langle F(\hat\varphi) \rangle _{t_1}=tr (F(\hat\varphi) \hat\rho(t_1))\\
    =\int \mathfrak{ D}\xi(\vec x)\  F(\xi) \langle \xi|\hat U(t_1,t_0) \hat\rho(t_0) \hat U(t_0,t_1)|\xi \rangle,
%  = \int \mathfrak{ D}\xi \int \mathfrak{ D}\xi_1 \int \mathfrak{ D}\xi_2\  F(\xi) \langle\xi|\hat U(t_1,t_0)|\xi_1\rangle
%  \langle\xi_1|\hat\rho(t_0)|\xi_2\rangle
%\langle\xi_2| \hat U(t_0,t_1)|\xi\rangle.
\label{genev}
\end{multline}
where evolution of the density matrix $\hat\rho(t)$ is governed by the evolution operator $\hat U(t,t_0)$
\begin{equation}
 \hat\rho(t)=\hat U(t,t_0) \hat\rho(t_0) \hat U(t_0,t),
\end{equation}
  $|\xi\rangle$ is an eigenstate of the field operator
  \mbox{$\hat \varphi(\vec x) |\xi\rangle = \xi(\vec x)|\xi\rangle$}   and
 $\int \mathfrak{D}\xi(\vec x)$ is a path integral over all possible functions
 $\xi(\vec x)$ originating from
 unity operator \mbox{$\hat 1 =\int \mathfrak{D}\xi(\vec x)\ |\xi\rangle \langle\xi|$.}

   After the usual procedure of the unity operator insertion we obtain
	 the matrix elements of the evolution operator which
	  path-integral representation is
		\footnote{We denote path integrals over space and time functions as
		$\int\Df f(t,\vec x) $,
		 whereas integrals over functions constant in time as
		 $\int \mathfrak{D}f(\vec x)$}
    \begin{gather*}
  \langle \xi|\hat U(t_1,t_0)|\xi_1 \rangle
    = \int\limits_{\eta_F(t_0,\vec x)=\xi_1(\vec x)}^{\eta_F(t_1,\vec x)
  =\xi(\vec x)} \Df \eta_F(t,\vec x) e^{i S[\eta_F]},\\
\langle\xi_2|\hat U(t_0,t_1)|\xi\rangle = \int\limits_{\eta_B(t_0,\vec x)=\xi_2(\vec x)}^{\eta_B(t_1,\vec x)
 =\xi(\vec x)} \Df \eta_B(t,\vec x) e^{-i S[\eta_B]}.
  \end{gather*}

Here $\eta_F(t,\vec x)$  and $\eta_B(t,\vec x)$  are the fields that lie on the  forward ($\eta_F$)
and backward ($\eta_B$) sides of the Keldysh contour (see \cite{LR1} for details).
Thus the observable \eqref{genev} reads
\begin{gather}
   \langle F(\hat\varphi) \rangle _{t_1}
  =   D[\xi_1,\rho_0,\xi_2] \int \mathfrak{ D}\xi\   F(\xi)\ \times\\
  \int\limits_{\eta_F(t_0,\vec x)=\xi_1(\vec x)}^{\eta_F(t_1,\vec x)
  =\xi(\vec x)} \Df \eta_F(t,\vec x)
	 \int\limits_{\eta_B(t_0,\vec x)=\xi_2(\vec x)}^{\eta_B(t_1,\vec x)
 =\xi(\vec x)} \Df \eta_B(t,\vec x)\ e^{iS_K[\eta_F,\eta_B]}\nonumber
\end{gather}
where integration over initial configuration and the Keldysh action are
\begin{gather*}
	D[\xi_1,\rho_0,\xi_2]=\int \mathfrak{ D}\xi_1 \int \mathfrak{ D}\xi_2\
	 \langle\xi_1|\hat\rho(t_0)|\xi_2\rangle,\\
	 S_K[\eta_F,\eta_B]=S[\eta_F]-S[\eta_B].
\end{gather*}

The final point of the trajectories which we integrate over is
the time of observation $t_1$. However, it is convenient to
extend the Keldysh contour to infinity so that the $t_1$
 remains only in the observable $F$.
The semiclassical decomposition is more evident with
the following change of variables
\footnote{One can meet equivalent notations
for such rotation in the literature
 $\phi_c\equiv \phi_r$ and $\phi_q \equiv \phi_a $ }
 (often called the Keldysh rotation)
\begin{equation}
\phi_c = \frac{\eta_F+\eta_B}{2}, \qquad \phi_q = \eta_F -\eta_B.
\end{equation}
Then general expression for the observable reads
\begin{gather}
    \langle F(\hat\varphi)\rangle_{t_1}
   =  D[\xi_1,\rho_0,\xi_2] \int \mathfrak{D}\chi
\int\limits_{\phi_c(t_0,\vec x)=\frac{\xi_1(\vec x)+\xi_2(\vec x)}{2}}^{\phi_c(\infty,\vec x) =\chi(\vec x)}  \Df \phi_c
	\nonumber   \\
\times
  \int\limits_{\phi_q(t_0,\vec x)=\xi_1(\vec x)-\xi_2(\vec x)}^{\phi_q(\infty,\vec x)=0}  \Df \phi_q
  \ F(\phi_c(t_1))
  \ e^{i S_K[\phi_c,\phi_q]} .
	\label{general2}
\end{gather}

This formula is rather general, hence we need to specify the Lagrangian. We use
 a scalar model with a quartic interaction term.
  \begin{gather}\label{static_lagr}
  \lagr = \frac{1}{2}\partial_{\mu}\varphi\partial^{\mu}\varphi - \frac{g^2}{4} \varphi^4 + J \varphi,\\
  S = \int d^4x \lagr. \nonumber
  \end{gather}
 Here $J(t,\vec x)$ is an auxiliary source which is kept
 to perform semiclassical decomposition.
  This source should be set to zero at the end of  calculations.

For the Lagrangian (\ref{static_lagr})  the Keldysh action
(after integration by parts) reads
\begin{gather}
 S_K[\phi_c,\phi_q] =
 \int d^3x\ \dot\phi_c(t_0,\vec x)(\xi_1(\vec x)-\xi_2(\vec x))\nonumber\\
 -
 \int d^3x \int\limits_{t_0}^{\infty}dt\ \Big(\phi_q A[\phi_c]
  -\frac{g^2 }{4}\phi_c\phi_q^3\Big), \\
 A[\phi_c]=[\partial_{\mu}\partial^{\mu}\phi_c+g^2\phi_c^3 - J].
 \label{static_action}
\end{gather}
Note that the term $\phi_q A[\phi_c]=0$ corresponds to projecting onto
the classical equation of motion for the Lagrangian \eqref{static_lagr}.

The semiclassical approximation of \eqref{static_action} means
expansion on $\phi_q$ around its saddle-point value

\begin{gather}
e^{ i\frac{g^2 }{4}\int\limits_{t_0}^{\infty}dt \int d^3 x\ \phi_c\phi_q^3}= \underbrace{1}_{LO}
 +\underbrace{\frac{ig^2 }{4}\int\limits_{t_0}^{\infty}dt \int d^3 x\ \phi_c\phi_q^3}_{NLO}+....
 \label{expansion}
 \end{gather}

This expansion does not require smallness of the coupling constant $g$.
Practically, the Leading Order contribution contains
quantum fluctuation up to one loop order.

{\bf The Leading Order} contribution to observables corresponds
to the first term in decomposition (\ref{expansion}).
The integration over $\phi_q$ and $\phi_c$ fields gives
(see \cite{LR2,method} for details)
 \begin{multline} \label{static_LO}
  \langle F(\hat\varphi)\rangle_{t_1}=\\
   = \int \mathfrak{D}\alpha(\vec x)  \mathfrak{D} p(\vec x) f_W[\alpha(\vec x),p(\vec x)),t_0] F(\phi_{cl}(t_1,\vec x)),
  \end{multline}
  where
  \begin{multline}
  f_W[\alpha(\vec x),p(\vec x),t_0] =\int \mathfrak{D}\beta(\vec x)
   \Big\langle\alpha+\frac{\beta}{2}\Big|\hat\rho(t_0)\Big|\alpha-\frac{\beta}{2}\Big\rangle \\
   \times \exp\Bigg(i \int  d^3 x\ p(\vec x) \beta(\vec x)  \Bigg).
   \end{multline}
	 is the Wigner functional defining initial state of the system,
$\phi_{cl}$ is the solution of classical equation of motion
    \begin{gather}
   \partial_{\mu}\partial^{\mu}\phi_{cl}+g^2\phi_{cl}^3 = J\Big|_{J=0}=0
	 \label{classEoM}
   \end{gather}
   with initial conditions given by
   \begin{gather}
   \phi_{cl}(t_0,\vec x) = \alpha(\vec x), \quad
   \dot\phi_{cl}(t_0,\vec x) = p(\vec x)
 \end{gather}
 and at zero axillary source $J(t,\vec x)$.

Let us introduce new notation for averaging over initial conditions
\begin{gather}\label{ic}
 \langle \mathcal{O}\rangle_{i.c.}
   = \int \mathfrak{D}\alpha(\vec x)  \mathfrak{D} p(\vec x)  f_W[\alpha(\vec x),p(\vec x)),t_0]\ \mathcal{O}.
\end{gather}
Then we can rewrite (\ref{static_LO}) shorter as
\begin{gather}
 \langle F(\hat\varphi)\rangle_{t_1}^{LO} =
 \langle F(\phi_{cl}(t_1,\vec x))  \rangle_{i.c.} .
\end{gather}

{\bf The Next-to-Leading Order} of the semiclassical decomposition
 (or quantum corrections to the CSA) is calculated as the second term of the
 expansion (\ref{expansion}).
 The path integration over $\phi_q$
can not be done as easy as at LO level because of the additional  $\phi_q^3$ part.
 However, each
$\phi_q$ can be replaced by functional derivative over source $J$
due to $\phi_q J $ term in the Keldysh action (\ref{static_action}) as
\begin{gather}
 \frac{\delta }{\delta J(t,\vec x)} e^{i S_K[\phi_c,\phi_q]} = i \phi_q(t,\vec x)e^{i S_K[\phi_c,\phi_q]}.
\end{gather}
This observation allows to perform functional integration
over $\phi_q$ and $\phi_c$ to obtain the answer for expectation value of the observable
up to NLO level
\begin{gather}\label{static_NLO}
 \langle F(\hat\varphi)\rangle_{t_1}^{LO+NLO} =
  \Bigg\langle F(\phi_{cl}(t_1,\vec x))\\
  + \frac{g^2}{4} \int\limits_{t_0}^{t_1}dt_2 \int d^3 x_2
  \phi_{cl}(t_2,\vec x_2)\frac{\delta^3 F(\phi_{cl}(t_1,\vec x))}{\delta J^3(t_2,\vec x_2)}\Bigg|_{J=0}  \Bigg\rangle_{i.c.}.\nonumber
\end{gather}
The expression above shows that there is no necessity in any new information
for evaluation of the NLO correction.
One should find the classical trajectory as a function of the initial
conditions, perform three variations over auxiliary source, integrate over
intermediate time and average with the Wigner functional. It is easy to recast
all terms of the semiclassical approximation to the following general form
\begin{gather}
	\label{all_terms}
 \langle F(\hat\varphi)\rangle_{t_1}=\\
 = \Bigg\langle\bar T e^{ \frac{g^2}{4}\displaystyle
  \int d\tau d\vec y\
  \phi_{cl}(\tau,\vec y)
  \frac{\delta^3}{\delta J^3(\tau,\vec y)}
  }\ F(\phi_{cl}(t_1,\vec x))\Bigg\rangle_{i.c.}\nonumber
\end{gather}
Here $\bar T$ denote the anti-time ordering which is required
to recover exponential form. The formula \eqref{all_terms} shows that
the building block of the semiclassical decomposition is the full
nonperturbative solution of the classical EoM $\phi_{cl}$ rather
than the  Green's function of the perturbative approach.
Hence, the strong field limit can be considered with the semiclassical
method, however,
 only for the narrow range of problems allowing
 the semiclassical decomposition itself.

 {\bf Numerical calculations} can be slightly  simplified.
Let us define $k$-th variation of the classical solution over source $J$ as
\begin{equation}
\frac{\delta^k\phi_{cl}(x_1)}{\delta J^k( x_2)}
= \Phi_k(x_1;  x_2).
\end{equation}
Then
\begin{multline}
 \frac{\delta^3 F(\phi_{cl}(x_1))}{\delta J^3( x_2)}
 = \frac{\partial F}{\partial \phi_{cl}}\Phi_3( x_1;  x_2)
 + \frac{\partial^3 F}{\partial \phi_{cl}^3} \Phi_1^3(x_1;  x_2)
 \\
 + 3 \frac{\partial^2 F}{\partial \phi_{cl}^2}
  \Phi_1(x_1;  x_2)\Phi_2( x_1; x_2) .
\end{multline}
Functions $\Phi_k( x_1; x_2)$ can be found by
variation of the classical EoM.
%\vskip -0.2cm
\begin{equation}
 \frac{\delta^3}{\delta J^3( x_2)}
 \left( \partial_{\mu}\partial^{\mu}\phi_{cl}( x_1)
 + g^2\phi_{cl}^3( x_1) = J( x_1) \right).\nonumber
\end{equation}
 Hence, to calculate the quantum correction to the CSA one need
 to find the solution
 of four linked differential equations
\begin{gather}
	\partial_{\mu}\partial^{\mu}\phi_{cl}( x_1)
 + g^2\phi_{cl}^3( x_1) =0,\nonumber\\
 L_{t_1}\Phi_1( x_1;x_2) = \delta^{(4)}( x_1 -x_2),\nonumber\\
 L_{t_1}\Phi_2( x_1; x_2) = -6 g^2 \phi_{cl}( x_1)\Phi_1^2( x_1; x_2),\nonumber\\
 L_{t_1}\Phi_3(x_1;x_2) = -6 g^2 \Phi_1^3( x_1;x_2)\nonumber\\
 - 18 g^2 \phi_{cl}( x_1)\Phi_1(x_1; x_2)\Phi_2( x_1;x_2)\nonumber\\
 L_{t_1} = \partial^2_{t_1} - \partial^2_{\vec x_1} + 3 g^2 \phi_{cl}^2( x_1).
\end{gather}
 without knowledge
of the exact dependence of the classical solution $\phi_{cl}(x)$ of
auxiliary source $J(x)$.

\end{document}